\documentclass{article}
\usepackage{amsmath}
\usepackage{footmisc}
\usepackage{graphicx}
\usepackage{amssymb}
\usepackage{enumerate}
\usepackage{multirow}
\usepackage{longtable}
\usepackage{booktabs}
\usepackage{multicol}
\usepackage{amsthm}
\begin{document}

\cleardoublepage \pagestyle{myheadings}
\bibliographystyle{plain}
\title{Alice-Bob systems, $P_s$-$T_d$-$C$ principles and multi-soliton solutions}
\author{S. Y. Lou$^{1,2}$\thanks{Email: lousenyue@nbu.edu.cn}  
\\
\footnotesize $^{1}$ \it Shanghai Key Laboratory of Trustworthy Computing, East China Normal University, Shanghai 200062, China,\\
\footnotesize $^{2}$\it Ningbo Collabrative Innovation Center of Nonlinear Harzard System of Ocean and Atmosphere\\
\footnotesize \it and Faculty of Science, Ningbo University,  Ningbo, 315211, China}
\maketitle

\begin{abstract}
To describe two-place physical problems, many possible models named Alice-Bob (AB) systems are proposed. To find and to solve these systems, the Parity (P), time reversal (T), charge conjugation (C), shifted-parity ($P_s$, parity with a shift), delayed time reversal ($T_d$, time reversal with a delay) and their possible combinations such as PT, PC, $P_sC$, $P_sT_d$ and $P_sT_dC$ etc. can be successively used. Especially, some special types of $P_s$-$T_d$-$C$ group invariant multi-soliton solutions for the KdV-KP-Toda type, mKdV-sG type, NLS type and discrete $H_1$ type AB systems are explicitly constructed.  \\
\bf PACS numbers: 
\rm  
05.45.Yv 
02.30.Ik 
11.30.Er 
11.10.Lm	
03.50.Kk 
02.30.Jr
02.30.Ks

\end{abstract}

\section{Introduction}
Usually, physical models are locally established around a single place (around a single space-time point), say, $\{x,\ t\}$. However, various physical problems may happen in two or more places which are linked each other, for instance, in quantum physics, two far-away particles (for simplicity, we call them Alice and Bob, respectively) may be entangled each other. Whence the state of Alice, $A(x,t)$, is given, the Bob's state, 
\begin{equation}
B(x',\ t')=\hat{f}A=A^f,
\end{equation}
for a suitable operator $\hat{f}$ may be determined at the same time. Usually, $\{x',\ t'\}$ is not neighbour to $\{x,\ t\}$. Thus, the intrinsic two-place models, the Alice-Bob systems (ABSs), are nonlocal. Some special types of this kind of nonlocal models have been proposed, such as the nonlocal nonlinear Schr\"odinger equations (NLS) \cite{AM} with $$\hat{f}=\hat{P}\hat{C}$$       
where $\hat{P}$ and $\hat{C}$ are the usual 
parity and charge conjugation operators. 

On the other hand, symmetries play an essential role in basic physical theories. For instance, the standard model is established by means of the $SU(3)\times SU(2) \times U(1)$ symmetry and all the predictions of the model have been completely proved. 18 Nobel prizes have been nominated  to the scientists who made great achievements on the model. Symmetries are not only the basic principles to establish physical models but also the key points to solve complicated scientific problems. The parity, time reversal ($\hat{T}$), charge conjugation, space and time translations and their suitable combinations (say $\hat{P}\hat{C}\hat{T}$ in weak interaction processes) are conserved for most of basic physical problems. Therefore, almost all the important nonlinear models possess these (or some of these) symmetries. Though these symmetries exist in various physical models, they, especially the $\hat{P}$-$\hat{C}$-$\hat{T}$ (parity and/or charge-conjugate and/or time-reversal) symmetries, are not directly used to solve nonlinear physical systems. 

In this paper, we will reveal that these symmetries are not only very useful to established some types of ABSs but also quite effective to find exact $\hat{P}$-$\hat{T}$-$\hat{C}$ invariant solutions of ABSs.      
In section 2, the ABSs are defined in detail by introducing the AB-BA principle  (AB-BA equivalence principle) and the $\hat{P}_s$-$\hat{T}_d$-$\hat{C}$ principle (shifted parity ($\hat{P}_s$) and/or delayed time reversal ($\hat{T}_d$) and/or charge conjugate invariance principle).  
Especially, various particular integrable ABSs, such as the AB-KdV systems, AB-MKdV systems, AB-KP systems, AB-NLS systems (the so-called nonlocal NLS models are included as special cases), AB-sine Gordon systems, AB-Toda systems and AB-$H_1$ systems are listed. In section 3, using the $\hat{P}_s$-$\hat{T}_d$-$\hat{C}$ principle, the $\hat{P}_s$-$\hat{T}_d$-$\hat{C}$ invariant multiple soliton 
solutions are explicitly obtained for various ABSs. 
The last section is a short summary and discussion.

\section{AB systems}
\subsection{AB-BA equivalence principle (AB-BA principle).} \rm It is reasonable that Alice and Bob are in the equivalent situation which implies  that Bob's coordinate system 
$x'\equiv \{x'_i,\ i=1,\ 2,\ \ldots,\ n\}$ should be 
linked with that of Alice 
$x\equiv \{x_i,\ i=1,\ 2,\ \ldots,\ n\}$ as 
\begin{equation}
x'=f(x)\equiv \hat{f}x.\label{Eqx'}
\end{equation} 
Inversely, Alice's coordinate system is related to that of Bob's by 
\begin{equation}
x=f^{-1}(x')\equiv \hat{f}^{-1}x'.\label{Eqx}
\end{equation} 
Our equivalence principle requires that the function $f$ and its inverse $f^{-1}$ should have the same form. In other words, 
\begin{equation}
\hat{f}^2=1. \label{Eqf}
\end{equation}
It is clear that there are infinitely many solutions of \eqref{Eqf}, since any solution of 
\begin{equation}
F(x',\ x)\pm F(x,\ x')=0 \label{Gf}
\end{equation}
solves \eqref{Eqf} for an arbitrary $F$. However, if the spacial relation \eqref{Eqx'} (and then \eqref{Eqx}) is further required to be only a linear transformation, then the solutions of \eqref{Eqf} can be completely fixed. 

For one dimensional transformation, there are only two linear transformation solutions of \eqref{Eqf},
the identity transformation 
\begin{equation}
x'=x \ (\mbox{for \ space} \ x), t'=t \ (\mbox{for \ time} \ t),  \label{1d1}
\end{equation}
and the shifted parity ($\hat{P}_s$)  transformation for the spatial variable,
 the delayed time reversal ($\hat{T}_d$) for the time variable
\begin{equation}
x'=-x+x_0\equiv \hat{P}_sx ,\quad t'=-t+t_0\equiv \hat{T}_dt. \label{1d2}
\end{equation}
When $x_0=0\ (t_0=0)$, $\hat{P}_s$ ($\hat{T}_d$) is reduced back to the pure parity $\hat{P}$ (time reversal $\hat{T}$). 

In two dimensional case, the allowed linear transformations of \eqref{Eqf} possess three cases,\\
the identity transformation 
\begin{equation}
\left(\begin{array}{l} x'_1\\ x'_2\end{array}\right) =\left(\begin{array}{l} x_1\\ x_2\end{array}\right)=\left(\begin{array}{ll} 1 & 0\\ 0 & 1\end{array}\right)\left(\begin{array}{l} x_1\\ x_2\end{array}\right), \label{2d1}
\end{equation}
the shifted parity with arbitrary constants $x_{10}$ and $x_{20}$
\begin{equation}
\left(\begin{array}{l} x'_1\\ x'_2\end{array}\right) =\left(\begin{array}{cc} -1 & 0\\ 0 & -1\end{array}\right)\left(\begin{array}{l} x_1\\ x_2\end{array}\right)+\left(\begin{array}{l} x_{10}\\ x_{20}\end{array}\right)\equiv \hat{P}_s^{x_1x_2}x, \label{2d2}
\end{equation} 
and the mixed shifted parity and rotation
\begin{equation}
\left(\begin{array}{l} x'_{i_1}\\ x'_{i_2}\end{array}\right) =\left(\begin{array}{cc} a_1a_2-1 & a_1\\ a_2(2-a_1a_2) & 1-a_1a_2\end{array}\right)\left(\begin{array}{l} x_{i_1}\\ x_{i_2}\end{array}\right)-\left(\begin{array}{l} -x_{10}\\ a_2x_{10}\end{array}\right), \label{2d3}
\end{equation}
where $\{i_1,\ i_2\}=\{1,\ 2\}$ or $\{2,\ 1\}$ and $\{a_1,\ a_2,\ x_{10}\}$ are arbitrary constants.  

In three dimensional case, we have four cases,\\
the identity transformation 
\begin{equation}
\left(\begin{array}{l} x'_1\\ x'_2 \\ x'_3 \end{array}\right) =\left(\begin{array}{l} x_1\\ x_2\\ x_3\end{array}\right)=\left(\begin{array}{ccc} 1 & 0 & 0\\ 0 & 1 & 0\\ 0 & 0& 1\end{array}\right)\left(\begin{array}{l} x_1\\ x_2\\ x_3 \end{array}\right), \label{3d1}
\end{equation}
the shifted parity with arbitrary constants $x_{10},\ x_{20}$ and $x_{30}$
\begin{equation}
\left(\begin{array}{l} x'_1\\ x'_2\\  x'_3 \end{array}\right) =\left(\begin{array}{ccc} -1 & 0 & 0\\ 0 & -1 & 0\\
0 & 0 & -1\end{array}\right)\left(\begin{array}{l} x_1\\ x_2\\ x_3\end{array}\right)+\left(\begin{array}{l} x_{10}\\ x_{20}\\ x_{30}\end{array}\right)\equiv \hat{P}_s^{x_1x_2x_3}x, \label{3d2}
\end{equation} 
the first type of mixed shifted parity and rotation ($A_1\equiv -a_1a_3-a_2a_4$)
\begin{equation}
\left(\begin{array}{l} x'_{i_1}\\ x'_{i_2}\\ x'_{i_3}\end{array}\right) =\left(\begin{array}{ccc} A_1-1 & a_2 & a_3\\ a_4(A_1-2) & 1+a_2a_4 & a_3a_4\\
a_1(A_1-2) & a_1a_2 & 1+a_1a_3\end{array}\right)\left(\begin{array}{l} x_{i_1}\\ x_{i_2} \\ x_{i_3}\end{array}\right)+\left(\begin{array}{c} x_{10}\\ a_4x_{10}\\ a_1x_{10}\end{array}\right), \label{3d3}
\end{equation}
and the second type of mixed shifted parity and rotation 
\begin{equation}
\left(\begin{array}{l} x'_{i_1}\\ x'_{i_2}\\ x'_{i_3}\end{array}\right) =\left(\begin{array}{ccc} a_1a_3-1 & a_2a_3 & a_3\\ a_1a_4 & a_2a_4-1 & a_4\\
a_1(A_1+2) & a_2(A_1+2) & A_1+1\end{array}\right)\left(\begin{array}{l} x_{i_1}\\ x_{i_2} \\ x_{i_3}\end{array}\right)+\left(\begin{array}{c} -x_{10}\\ -x_{20}\\ a_1x_{10}+a_2x_{20}\end{array}\right). \label{3d4}
\end{equation}
In \eqref{3d3} and \eqref{3d4}, $\{i_1,\ i_2,\ i_3\}$ are any permutations of $\{1,\ 2,\ 3\}$, i.e, $\{i_1,\ i_2,\ i_3\}=\{1,\ 2,\ 3\},$ $ \{1,\ 3,\ 2\},$ $ \{2,\ 1,\ 3\},$ $ \{2,\ 3,\ 1\},$ $ \{3,\ 1,\ 2\} $ or $\{3,\ 2,\ 1\}$, and $\{a_1,\ a_2,\ a_3,\ a_4,\ x_{10},\ x_{20}\}$ are arbitrary constants. 

Using the above AB-BA equivalence principle, our most general AB systems have the form
\begin{eqnarray}
&&CAB\equiv \left\{ \begin{array}{l}
G_1(A,\ B)=0,\\
G_2(A,\ B)=0,
\end{array}\right.\label{G}\\
&&B=A^f=\hat{f}A \label{AB}
\end{eqnarray}
with the AB-BA equivalence condition
\begin{equation}
\hat{f}^2=1,\ x'=\hat{f}x,\ x=\hat{f}x' \label{ABBA}
\end{equation}
for a suitable $\hat{f}$ as listed in \eqref{1d2}, \eqref{2d2},  \eqref{2d3}, \eqref{3d2}, \eqref{3d3}, \eqref{3d4} and so on. 

Obviously, an ABS is a coupled system $CAB$ \eqref{G} which is consistent under the correlated condition \eqref{AB}. 

Because of the condition \eqref{ABBA}, the general coupled system \eqref{G} should be invariant under the transformation $\hat{f}$, i.e.,
\begin{equation}
\hat{f}CAB = CAB.
 \label{fCAB}
\end{equation}

For physically significance,  we only discuss the coupled system \eqref{G} known in natural science.

\subsection{AB-KdV systems}
In nonlinear physics, the KdV equation ($A=A(x,t)$) \cite{KdV}
\begin{equation}
KdV\equiv A_t+A_{xxx}+6AA_x=0 \label{KdV}
\end{equation}
is one of the most important and famous models which can be applied in almost all  physical fields especially in fluid related branches \cite{KdVa}. 

The most general coupled homogeneous KdV system may have the form
\begin{eqnarray}
CKdV&\equiv &\left\{\begin{array}{ll}
A_t+a_1A_{xxx}-a_2B^f_{x'x'x'}
+(a_3A+a_4B^f)A_x-(a_5A+a_6B^f)B^f_{x'}=0,
\\ \\
B^f_{t'}-b_1A_{xxx}+b_2B^f_{x'x'x'}
-(b_3A+b_4B^f)A_x+(b_5A+b_6B^f)B^f_{x'}=0,
\end{array} \right.\label{ABKa}  \\
&=&\left\{\begin{array}{ll}
A_t+a_1A_{xxx}-a_2B^f_{x'x'x'}
+(a_3A+a_4B^f)A_x-(a_5A+a_6B^f)B^f_{x'}=0,
\\ \\
B_{t}-b_1A^f_{x'x'x'}+b_2B_{xxx}
-(b_3A^f+b_4B)A^f_{x'}+(b_5A^f+b_6B)B_{x}=0,
\end{array} \right.\label{ABKb} \\
&=&\left\{\begin{array}{ll}
A_t+a_1A_{xxx}+a_2B_{1xxx}
+(a_3A+a_4B_1)A_x+(a_6A+a_7B_1)B_{1x}=0,
\\ \\
B_{1t}+b_1A_{xxx}+b_2B_{1xxx}
+(b_3A+b_4B_1)A_{x}+(b_5A+b_6B_1)B_{1x}=0,
\end{array} \right. 
 \label{ABKc} 
\end{eqnarray}
where 
\begin{eqnarray}
&&A=A(x,\ t),\ A^f= \hat{f}A=A(x',\ t'),\ B=B(x,\ t),\ B^f= \hat{f}B=B(x',\ t') =B_1(x,\ t),\nonumber\\
&& 
x'=\hat{f}x =-x+x_0,\ t'=\hat{f}t =-t+t_0, \ \hat{f}=\hat{P}_s\hat{T}_d. \label{fPT}
\end{eqnarray}
The general coupled KdV system 
possesses three equivalent expressions \eqref{ABKa}, \eqref{ABKb} and \eqref{ABKc}. However, the physical meanings of them are quite different. Equation \eqref{ABKc} is a usual single place nonlinear model, \eqref{ABKa} describes the two-place nonlinear interaction between $A(x,\ t)$ and $B(x',\ t')$, while \eqref{ABKb} displays an interaction model between two-place quantities $\{A(x,\ t),\ B(x, t)\}$ and 
$\{A(x'\ t'),\ B(x', t')\}$. 

The general coupled one-place KdV system \eqref{ABKc} has been studied in many literatures \cite{CKdV,CKdVa,CKdVb,
CKdVc,CKdVd,WDS,Fordy1}. Its possible integrable classification has been done by the general symmetry analysis, bi-Hamiltonian structure, Painlev\'e test and prolongation search on its Lax pairs. The model can also been derived from two layer fluids \cite{CKdV}. 

Due to the equivalence of \eqref{ABKa}, \eqref{ABKb} and \eqref{ABKc}, it may be more interesting to find some possible two-place reductions which are not equivalent to 
some one-place models. 

It is fortunate that the special case of \eqref{ABKb}
\begin{eqnarray}
SCKdV&\equiv &\left\{\begin{array}{ll}
A_t+a_1A_{xxx}-a_2B^f_{x'x'x'}
+(a_3A+a_4B^f)A_x-(a_6A+a_7B^f)B^f_{x'}=0,
\\ \\
B_{t}-a_2A^f_{x'x'x'}+a_1B_{xxx}
-(a_7A^f+a_6B)A^f_{x'}+(a_4A^f+a_3B)B_{x}=0,
\end{array} \right.\label{ABKbs} 
\end{eqnarray} 
allows an intrinsic two-place AB-KdV reduction  
\begin{eqnarray}
&&ABKdV\equiv 
A_t+a_1A_{xxx}+a_2B_{xxx}
+(a_3A+a_4A^f)A_x+(a_6A+a_7A^f)B_{x}=0,
\label{AAKbs} 
\\
&& B=A^f=A^{P_sT_d}
=A(-x+x_0,\ -t+t_0), \nonumber
\end{eqnarray} 
by directly taking $B=A$  from \eqref{ABKbs}.

Here we list four special integrable realizations of \eqref{AAKbs}.

The first type of integrable AB-KdV reduction has the form  
\begin{eqnarray}
ABKdV_1 \equiv  
A_t-\frac12A_{xxx}+\frac32B_{xxx}
-3(A-B)A_x+6AB_{x}=0,\quad B= A^{{P}_s{T}_d}.
\label{AKdV1}
\end{eqnarray}
It is integrable because it can also be obtained from the reduction of the Hirota-Satsuma system \cite{HS} and thus one can obtain its Lax pair \cite{Fordy}
\begin{equation}
\left(\begin{array}{c}
\psi_1 \\ \psi_2
\end{array}\right)_{xx}=\left(\begin{array}{cc}
-A & \lambda_1\\ \lambda_2 & -B
\end{array}\right)\left(\begin{array}{c}
\psi_1 \\ \psi_2
\end{array}\right),
\end{equation}
\begin{equation}
\left(\begin{array}{c}
\psi_1 \\ \psi_2
\end{array}\right)_{t}=\left(\begin{array}{cc}
\frac12(3B-A)_{x}+(A-3B)\partial_x & -4\lambda_1\partial_x\\ -4\lambda_2\partial_x & \frac12(3A-B)_x+(B-3A)\partial_x
\end{array}\right)\left(\begin{array}{c}
\psi_1 \\ \psi_2
\end{array}\right)
\end{equation}
with two arbitrary spectral parameters $\lambda_1$ and $\lambda_2$. 

The second type of AB-KdV has the form  
\begin{eqnarray}
ABKdV_2 \equiv  
A_t+A_{xxx}+3(A+B)A_x=0,\quad B=\hat{P}_s\hat{T}_d A.
\label{AKdV2}
\end{eqnarray}
When $\{x',\ t'\}$ is changed as $\{x,\ t\}$, 
the $ABKdV_2$ is reduced back to the known KdV equation. 
It is not difficult to prove that the AB-KdV model \eqref{AKdV2} is  integrable with the Lax pair
\begin{eqnarray}
&&\left(\begin{array}{c}
\psi_1\\ \psi_2\end{array}\right)_{xx}=\left(\begin{array}{cc}
-\frac12 C-\lambda_1 & 0\\ \lambda_2-\lambda_4A_x-\lambda_5B_x+(\lambda_3+\lambda_6C)\partial_x & -\frac12C-\lambda_1\end{array}\right)\left(\begin{array}{c}
\psi_1\\ \psi_2 \end{array}\right)
,\label{Lt1}\\
&&\left(\begin{array}{c}
\psi_1\\ \psi_2\end{array}\right)_t=\left(\begin{array}{cc}
\frac12C_x+\lambda_8+(4\lambda_1-C)\partial_x &  0\\ E-D\partial_x & 
\frac12 C_x+\lambda_8+(4\lambda_1-C)\partial_x \end{array}\right)\left(\begin{array}{c}
\psi_1\\ \psi_2 \end{array}\right), \nonumber
\end{eqnarray}
where 
\begin{eqnarray}
&& C\equiv A+B,\nonumber\\ 
&& D\equiv \lambda_6C_x+2(\lambda_4A_x+\lambda_5B_x)+4\lambda_2, \nonumber
\\
&& E\equiv \lambda_7+
\lambda_4A_{xx}+\lambda_5B_{xx}
+\frac12 \lambda_3C-\frac14 C\lambda_6(C+8\lambda_1).
\end{eqnarray}
The third type of integrable AB-KdV system possesses the form
\begin{equation}
ABKdV_3\equiv A_t+A_{xxx}+\frac34\left[(A+B)^2\right]_x=0,\quad B=A(-x+x_0,\ -t+t_0) \label{AKdV3}
\end{equation} 
with the following Lax 
pair
\begin{eqnarray}
&&\left(\begin{array}{c}
\psi_1\\ \psi_2\end{array}\right)_{xx}=\left(\begin{array}{cc}
-\frac12 C-\lambda_1 & 0\\ \lambda_2-C_{1x}+(\lambda_6+\lambda_5 C-2C_{1})\partial_x & -\frac12C-\lambda_1\end{array}\right)\left(\begin{array}{c}
\psi_1\\ \psi_2 \end{array}\right)
,\label{Lt3}\\
&&\left(\begin{array}{c}
\psi_1\\ \psi_2\end{array}\right)_t=\left(\begin{array}{cc}
\frac12C_{x}+\lambda_8+(4\lambda_1-C)\partial_x &  0\\ E_1-(4\lambda_2+\lambda_5C_x)\partial_x & 
\frac12 C_x+\lambda_8+(4\lambda_1-C)\partial_x \end{array}\right)\left(\begin{array}{c}
\psi_1\\ \psi_2 \end{array}\right), \nonumber
\end{eqnarray} 
where 
\begin{eqnarray}
&&C\equiv A+B,\quad C_1\equiv (\lambda_3A+\lambda_4B),\nonumber\\
&& E_1\equiv C_{1xx}+4\lambda_1C_1+\frac12\lambda_6C+\lambda_7
-\frac14C\left[\lambda_5(C+8\lambda_1)+C_1-3\lambda_4A
-3\lambda_3B\right] .
\end{eqnarray}
The fourth type of the AB-KdV system can be written as
\begin{equation}
ABKdV_4\equiv A_t+A_{xxx}+\frac32(B+2A)A_x+\frac32AB_x=0,\ B=A^{P_sT_d}. \label{AKdV4}
\end{equation}
The integrability of the fourth AB-KdV \eqref{AKdV4} is guaranteed by the following Lax pair ($C_2\equiv \lambda_3A+\lambda_2B $)
\begin{eqnarray}
&&\left(\begin{array}{c}
\psi_1\\ \psi_2\end{array}\right)_{xx}=\left(\begin{array}{cc}
-\frac12 C-\frac14 \lambda_1 & 0\\ -C_{2x}-C_2 \partial_x & -\frac14\lambda_1\end{array}\right)\left(\begin{array}{c}
\psi_1\\ \psi_2 \end{array}\right)
,\label{Lt4}\\
&&\left(\begin{array}{c}
\psi_1\\ \psi_2\end{array}\right)_t=\left(\begin{array}{cc}
\frac12C_{x}+\lambda_5+(\lambda_1-C)\partial_x &  0\\ C_{2xx}+\frac12 C_2 (\lambda_1+2C)-C_{2x}\partial_x & 
\lambda_5+\lambda_1\partial_x \end{array}\right)\left(\begin{array}{c}
\psi_1\\ \psi_2 \end{array}\right). \nonumber
\end{eqnarray} 
It is also interesting that, the fourth type of the AB-KdV system \eqref{AKdV4} possesses another Lax pair:
\begin{eqnarray}
&&\left(\begin{array}{c}
\psi_1\\ \psi_2\end{array}\right)_{xx}=\left(\begin{array}{cc}
-\frac12 C-\lambda_1 & 0\\ \lambda_2-C_{1x}+(\lambda_6+\lambda_5 C-2C_{1})\partial_x & -\frac12C-\lambda_1\end{array}\right)\left(\begin{array}{c}
\psi_1\\ \psi_2 \end{array}\right)
,\label{Lt41}\\
&&\left(\begin{array}{c}
\psi_1\\ \psi_2\end{array}\right)_t=\left(\begin{array}{cc}
\frac12C_{x}+\lambda_8+(4\lambda_1-C)\partial_x &  0\\ E_2-(4\lambda_2+\lambda_5C_x)\partial_x & 
\frac12 C_x+\lambda_8+(4\lambda_1-C)\partial_x \end{array}\right)\left(\begin{array}{c}
\psi_1\\ \psi_2 \end{array}\right), \nonumber
\end{eqnarray} 
with 
\begin{equation}
E_2\equiv C_{1xx}+4\lambda_1C_1+\frac12\lambda_6C-\frac14C\left[\lambda_5(C+8\lambda_1)-2C_1\right]+\lambda_7.
\end{equation}

Finally, it is worth to emphasize that all the AB-KdV systems listed above possess the $\hat{P}_s\hat{T}_d=\hat{f}$ invariance and this invariance originates from the invariance of the coupled KdV \eqref{ABKc}
\begin{equation}
\hat{P}_s\hat{T}_d CKdV=CKdV.
\end{equation}

\subsection{AB-NLS systems}
 Another ubiquitous nonlinear physical model is the so-called nonlinear Schr\"odinger (NLS) equation 
 \begin{equation}
NLS\equiv iA_t+A_{xx}
+2\sigma |A|^2A=0,\ \quad \sigma=\pm1. \label{NLS}
 \end{equation}
Similar to the AB-KdV systems as discussed in the last subsection, to extend the NLS equation to possible two-place AB models, we rewrite it as a special  reduction of the AKNS system 
\begin{eqnarray}
AKNS&\equiv& \left\{ \begin{array}{l} 
iA_t +A_{xx}+2\sigma A^2B_1=0,\\
-iB_{1t} +B_{1xx}+2\sigma B_1^2A=0,\end{array}\right.
\label{AKNSa} \\
&=& \left\{ \begin{array}{ll} iA_t +A_{xx}+2\sigma A^2B^{P_s}=0, & (B_1=B^{P_s}), \\
-iB_t +B_{xx}+2\sigma B^2A^{P_s}=0, & \end{array}\right.
\label{AKNSb}\\
&=& \left\{ \begin{array}{ll} iA_t +A_{xx}+2\sigma A^2B^{P_ST_dC}=0, & (B_1=B^{P_sT_dC}),\\
-iB_t +B_{xx}+2\sigma B^2A^{P_sT_dC}=0.& \end{array}\right.
\label{AKNSc}
\end{eqnarray}
Similarly, though the systems \eqref{AKNSa}, \eqref{AKNSb}, and \eqref{AKNSc} are equivalent, their physical meanings are quite different. Especially, their reductions corresponding to $B=A^C=A^*$ (or $B_1=A^C=A^*$ for \eqref{AKNSa}) are completely different. The reduction of \eqref{AKNSa} is just the well known NLS equation \eqref{NLS}.
The reduction of \eqref{AKNSb} has the form
\begin{equation}
ABNLS_1\equiv iA_t+A_{xx}+2\sigma A^2B=0,\ 
B=A^{P_sC}= \hat{P}_s\hat{C}A=A^*(-x+x_0,\ t). \label{ABNLS1}
\end{equation}
When $x_0=0$, the $ABNLS_1$ \eqref{ABNLS1} is just the so-called nonlocal NLS which was first introduced by Ablowitz and Musslimani \cite{AM}. The model \eqref{ABNLS1} without $x_0$ has been widely studied by many authors \cite{PRE}. However, it is still a new model for $x_0\neq 0$. 

The reduction of \eqref{AKNSc} has the form 
\begin{equation}
ABNLS_2\equiv iA_t+A_{xx}+2\sigma A^2B=0,\quad B=A^{P_sT_d}= \hat{P}_s\hat{T}_dA=A(-x+x_0,\ -t+t_0).
\label{ABNLS2} 
\end{equation}
In our knowledge, the AB-NLS model \eqref{ABNLS2} is also new. It is a really two-place nonlocal model 
though it is reduced from a one-place coupled system. 

All the models in this subsection are $P_s$ invariant, $T_dC$ invariant and $P_sT_dC$ invariant.

The integrability of the AB-NLS systems \eqref{ABNLS1} and \eqref{ABNLS2} is guaranteed by the integrability of the AKNS system \eqref{AKNSa} with the 
well known Lax pair
\begin{equation}
\left(\begin{array}{c}
\psi_1\\ \psi_2
\end{array}\right)_x=\left(\begin{array}{cc}
\lambda & A \\ -\sigma B & -\lambda
\end{array}\right)
\left(\begin{array}{c}
\psi_1\\ \psi_2
\end{array}\right),
\end{equation}
\begin{equation}
\left(\begin{array}{c}
\psi_1\\ \psi_2
\end{array}\right)_t=\left(\begin{array}{cc}
2i\lambda^2+i\sigma AB & 2i\lambda A+iA_x \\ -i\sigma(2\lambda B+B_x) & -2i\lambda^2-i\sigma AB
\end{array}\right)
\left(\begin{array}{c}
\psi_1\\ \psi_2
\end{array}\right),
\end{equation}
where $B=A^{P_sC}=A^*(-x+x_0,t)$ for $ABNLS_1$ \eqref{ABNLS1} and $B=A^{P_sT_d}=A(-x+x_0,-t+t_0)$ for $ABNLS_2$ \eqref{ABNLS2}.

\subsection{AB-MKdV systems}
The well known modified KdV equation (MKdV) can be written in the form 
\begin{equation}
MKdV\equiv A_t+A_{xxx}+6\sigma A^2A_x=0,\quad \sigma=\pm1, \label{MKdV}
\end{equation}
which is a natural reduction of the third order AKNS system,
\begin{eqnarray}
AKNS3\equiv \left\{ \begin{array}{l}
A_t+A_{xxx}+6AB_1A_x=0,\\
B_{1t}+B_{1xxx}+6AB_1B_{1x}=0,
\end{array}\right. \label{AKNS3a}
\end{eqnarray}
by taking 
$$B_1=\sigma A.$$

It is clear that the one-place coupled model \eqref{AKNS3a} possesses an equivalent two-place form 
\begin{eqnarray}
AKNS3 \equiv \left\{
\begin{array}{l}
A_t+A_{xxx}+6AB^{f}A_x=0,\quad B^{f}\equiv B(-x+x_0,\ -t+t_0)\\
B_{t}+B_{xxx}+6A^{f}BB_{x}=0,
\end{array}\right.\label{AKNS3b}
\end{eqnarray}
by taking 
$$B_1=B^{f}.$$

Starting from the two-place form \eqref{AKNS3b}, one can readily obtain an intrinsic two-place AB-MKdV equation 
\begin{equation}
ABMKdV\equiv A_t+A_{xxx}+6\sigma ABA_x=0,\quad B=A^f=\hat{P}_s\hat{T}_dA=A(-x+x_0,\ -t+t_0). \label{ABMKdV}
\end{equation}
 The model \eqref{ABMKdV} is also a new model though a special case of \eqref{ABMKdV} with $x_0=0$ and $t_0=0$ has been given by Ji and Zhu \cite{Zhu}. 

The ABMKdV \eqref{ABMKdV} is integrable because it is a reduction of the third order 
AKNS system. Thus its Lax pair can be obtained from that of the AKNS system,
\begin{equation}
\left(\begin{array}{c}
\psi_1\\ \psi_2
\end{array}\right)_x=\left(\begin{array}{cc}
-\lambda & A \\  -\sigma B & \lambda
\end{array}\right)
\left(\begin{array}{c}
\psi_1\\ \psi_2
\end{array}\right)
\end{equation}
\begin{equation}
\left(\begin{array}{c}
\psi_1\\ \psi_2
\end{array}\right)_t=\left(\begin{array}{cc}
4\lambda^3+2\sigma \lambda AB+\sigma AB_x-\sigma B A_x & -4\lambda^2 A+2\lambda A_x-A_{xx}-2\sigma A^2B
  \\ 4\sigma\lambda^2 B+2\sigma \lambda B_x+\sigma B_{xx}+2B^2A  & -4\lambda^3-2\sigma\lambda AB-\sigma AB_x+BA_x
\end{array}\right)
\left(\begin{array}{c}
\psi_1\\ \psi_2
\end{array}\right).
\end{equation}

Many other types of intrinsic AB-MKdV system can be found from the coupled MKdV systems, say, those obtained from suitable Miura transforms of the AB-KdV systems \cite{WDS}. 
 
\subsection{AB-sine-Gordon equation}
The well known sine-Gordon (sG) equation 
\begin{equation}
sG\equiv A_{xt}=\sin(A) \label{sG}
\end{equation}
is one of the most important integrable relativistic models which is widely used in the field theory, optics and condense matter physics. 
The sG model is also a special member of the AKNS hierarchy. Its general coupled AKNS form can be written 
as ($\sigma^2=1$)
\begin{eqnarray}
ABsG\equiv \left\{
\begin{array}{l}
A_{xt}=\left(\sigma\sqrt{a-\sin(A)\sin(B)}+A_xA_t\right)\tan(A)+\frac12 \frac{\tan(A)B_t+\tan(B)A_t}{\tan(A)\tan(B)-a\sec(A)\sec(B)}A_x,\\
B_{xt}=\left(\sigma\sqrt{a-\sin(A)\sin(B)}+B_xB_t\right)\tan(B)+\frac12 \frac{\tan(A)B_t+\tan(B)A_t}{\tan(A)\tan(B)-a\sec(A)\sec(B)}B_x,
\end{array}
\right.\label{ABsGa}
\end{eqnarray}
Though the coupled sG system \eqref{ABsGa} is one member of the AKNS hierarchy, it seems that the model have not yet been explicitly written down elsewhere. In addition to the known sG (or sinh-Gordon) reduction \eqref{sG} by taking $B=A$ and $a=1$ (or $B=A=i \phi$ and $a=1$), there exist some other simple interesting reductions which have not yet been studied. For instance, the tan-Gordon (TG) equation
\begin{equation}
A_{xt}=(c+A_xA_t)\tan(A),\label{TG}
\end{equation}
the tanh-Gordon (ThG) equation
\begin{equation}
A_{xt}=(c+A_xA_t)\tanh(A),\label{ThG}
\end{equation}
and the two-place model, the AB-sG model ($B\equiv A^{P_sT_d}=A(-x+x_0,\ -t+t_0)$),
\begin{equation}
A_{xt}=\left(\sigma\sqrt{a-\delta \sin(A)\sin(B)}+A_xA_t\right)\tan(A)+\frac12\frac{[\tan(A)B_t+\tan(B)A_t]A_x}{\tan(A)\tan(B)-a\sec(A)\sec(B)},\label{ABSG}
\end{equation}
can be reduced from \eqref{ABsGa} via 
$\{B=0,\ \sigma\sqrt{a}=c\}$, $\{B=0,\ A\rightarrow iA,\ \sigma\sqrt{a}=c\}$ and $\{B\rightarrow \delta A^{P_sT_d}, \ \delta^2=1\}$, respectively. 

The Lax pair of the coupled sG system \eqref{sG} and all their reductions can be written as 
\begin{eqnarray}
\left(\begin{array}{c}
\psi_1 \\ \psi_2
\end{array}\right)_x=\left(\begin{array}{cc}
-\lambda & \frac12 \frac{\sigma\cos(A)A_x}{\sqrt{a-\sin(A)\sin(B)}} \\  -\frac12 \frac{\sigma\cos(B)B_x}{\sqrt{a-\sin(A)\sin(B)}} & \lambda
\end{array}\right)\left(\begin{array}{c}
\psi_1 \\ \psi_2
\end{array}\right),
\end{eqnarray}
\begin{eqnarray}
\left(\begin{array}{c}
\psi_1 \\ \psi_2
\end{array}\right)_t=\frac1{4\lambda}\left(\begin{array}{cc}
-\sigma\sqrt{a-\sin(A)\sin(B)} &  \sin(A) \\  \sin(B) & \sigma\sqrt{a-\sin(A)\sin(B)}
\end{array}\right)\left(\begin{array}{c}
\psi_1 \\ \psi_2
\end{array}\right)
\end{eqnarray}
with the suitable selections of $B$ and $A$ for the corresponding models.

When $B=A$ and $a=1$, the AB-sG system \eqref{ABSG} returns back to the usual sG system \eqref{sG}. There are also many other types of AB-sG systems, for instance, the negative ones of the integrable AB-MKdV hierarchies. 

\subsection{AB-KP system}
In 2+1 dimensions, one of the most important nonlinear systems is the so-called KP equation 
\begin{equation}
KP\equiv A_{xt}=\left(\frac14A_{xxx}+6AA_x\right)_x+\frac34A_{yy}. \label{KP}
\end{equation}
In order to find some nontrivial AB-KP systems, one can rewrite some known coupled KP systems to their two-place forms. For instance, one of the coupled KP system
has the form 
\begin{eqnarray}
CKP&\equiv &\left\{
\begin{array}{l}
A_{xt}=\left(\frac14A_{xx}+3AB+\frac32A^2-\frac32B^2\right)_{xx}+\frac34A_{yy}, \\ \\
B_{xt}=\left(\frac14B_{xx}+3AB+\frac32B^2-\frac32A^2\right)_{xx}+\frac34B_{yy},
\end{array}
\right.\label{KPA}\\ &&\nonumber\\
&=&\left\{
\begin{array}{l}
A_{xt}=\left[\frac14A_{xx}+3AB^f+\frac32A^2-\frac32(B^f)^2\right]_{xx}+\frac34A_{yy},  \\ \\
B_{xt}=\left[\frac14B_{xx}+3A^fB+\frac32B^2-\frac32(A^f)^2\right]_{xx}+\frac34B_{yy},
\end{array}
\right.\label{KPB}
\end{eqnarray}
with
\begin{equation}
A^f=\hat{P}_s^{xy}\hat{T}_dA=A(-x+x_0,\ -y+y_0,\ -t+t_0),\label{Af1}
\end{equation}
or 
\begin{equation}
A^f=\hat{P}_s^x\hat{T}_dA=A(-x+x_0,\ y,\ -t+t_0), \label{Af2}
\end{equation}
or 
\begin{equation}
A^f=\hat{P}_s^yA=A(x,\ -y+y_0,\ t). \label{Af3}
\end{equation}

It is straightforward to see that the coupled KP system \eqref{KPB} permits an intrinsic two-place AB-KP reduction,
\begin{eqnarray}
ABKP&\equiv &
\begin{array}{l}
A_{xt}=\left[\frac14A_{xx}+3AB+\frac32A^2-\frac32B^2\right]_{xx}+\frac34A_{yy}, \quad B=A^f
\end{array}
\label{ABKP}
\end{eqnarray}
with $A^f$ given by \eqref{Af1}, \eqref{Af2} or \eqref{Af3}. 

The coupled KP system \eqref{KPA}
is Lax integrable with the Lax pair
\begin{equation}
\left(\begin{array}{c}
\psi_1 \\ \psi_2
\end{array}\right)_y = \left(\begin{array}{c}
\psi_1 \\ \psi_2
\end{array}\right)_{xx} + \left(\begin{array}{cc}
2(A+B) & -2(A-B) \\ 2(A-B) & 2(A+B)
\end{array}\right) \left(\begin{array}{c}
\psi_1 \\ \psi_2
\end{array}\right),\label{LxKP}
\end{equation} 
\begin{equation}
\left(\begin{array}{c}
\psi_1 \\ \psi_2
\end{array}\right)_t = \left(\begin{array}{c}
\psi_1 \\ \psi_2
\end{array}\right)_{xxx} + \left(\begin{array}{cc}
\alpha_+ & -\alpha_- \\ \alpha_- & \alpha_+
\end{array}\right) \left(\begin{array}{c}
\psi_1 \\ \psi_2
\end{array}\right),\label{LtKP}
\end{equation} 
where  
\begin{equation}
\alpha_{\pm}\equiv \frac32 (A\pm B)_x+ \frac32 \int (A\pm B)_y \mbox{\rm dx} +3(A\pm B)\partial_x. \label{alpha}
\end{equation}
For the AB-KP equation \eqref{ABKP}, the Lax pair possesses the same form as \eqref{LxKP} and \eqref{LtKP} but with $B=A^f$ and \eqref{Af1}, \eqref{Af2} or \eqref{Af3}. 

In fact, for the coupled KP system \eqref{KPA}, there exist some other complicated AB reductions if other AB-BA equivalent coordinates are taken as those shown in \eqref{2d3}, \eqref{3d3} and \eqref{3d4}.

\subsection{AB-Toda systems}
The AB systems can also be established in discrete cases. One of the most important discrete integrable model is the so-called 
Toda lattice which is equivalent to  
\begin{equation}
Toda\equiv (A_n+2)A_{n,xx}-A_{n,x}^2+\frac12(A_n+2)^2E^2A_{n-1}=0, \label{Toda}
\end{equation}
where the difference operator $E$ is defined as 
\begin{equation}
EA_n=A_{n+1}-A_n. \label{E}
\end{equation}

In order to find some intrinsic two-place models, the AB-Toda systems, we should rewrite the $Toda$ system \eqref{Toda} in the coupled forms. 

One of the simplest coupled Toda system may have the form 
\begin{eqnarray}
 CTD &\equiv &\left\{
\begin{array}{l}
[(A_n+2)^2+(B'_n+2)^2]A_{nxx}=(A_n+2)\left(A_{nx}^2-B^{'2}_{nx}\right)+2(B'_n+2)A_{nx}B'_{nx}
\\ 
\qquad \qquad +\frac14[(A_n+2)^2+(B'_n+2)^2]\left[(A_n+B'_n+4)E^2A_{n-1}+
(A_n-B'_n)E^2B'_{n-1}\right],\\ \\
\left\{(A_n+2)^2+(B'_n+2)^2\right\}
B'_{nxx}=(B'_n+2)\left(B^{'2}_{nx}-A_{nx}^2\right)+2(A_n+2)A_{nx}B'_{nx}
\\
\qquad \qquad +\frac{1}{4}
\left\{(A_n+2)^2+(B'_n+2)^2\right\}
\left[(A_n+B'_n+4)E^2B'_{n-1}
+(B'_n-A_n)E^2A_{n-1}\right],
\end{array}
\right. \label{CTDa}\\ \nonumber\\
&=& \left\{
\begin{array}{l}
[(A_n+2)^2+(B^f_n+2)^2]A_{nxx}=(A_n+2)\left[A_{nx}^2-(B^f_{nx})^2\right]+2(B^f_n+2)A_{nx}B^f_{nx}
\\
\qquad \qquad +\frac14[(A_n+2)^2+(B^f_n+2)^2]\left[(A_n+B^f_n+4)E^2A_{n-1}+
(A_n-B^f_n)E^2B^f_{n-1}\right],
\\ \\
\left[(A^f_n+2)^2+(B_n+2)^2\right]
B_{nxx}=(B_n+2)
\left[B_{nx}^2-(A^f_{nx})^2\right]
+2(A^f_n+2)A^f_{nx}B_{n,x}
\\
\qquad \qquad +\frac14[(A^f_n+2)^2+(B_n+2)^2]\left[(A^f_n+B_n+4)E^2B_{n-1}+
(B_n-A^f_n)E^2A^f_{n-1}
\right],
\end{array}
\right.\label{CTDb}
\end{eqnarray}
where
\begin{equation}
A_n^f=\hat{f}A_n=\hat{P}^x_s\hat{P}^n_sA_n=A(-x+x_0,-n+n_0),\label{na}
\end{equation}
or
\begin{equation}
A_n^f=\hat{f}A_n=\hat{P}^x_sA_n=A(-x+x_0,n),\label{nb}
\end{equation}
or
\begin{equation}
A_n^f=\hat{f}A_n=\hat{P}^n_sA_n=A(x,-n+n_0).\label{nc}
\end{equation}
It is interesting that in addition to the 
usual Toda reduction \eqref{Toda} from \eqref{CTDa} via $B'=A$, there are three 
intrinsic two-place Toda, AB-Toda reductions, 
\begin{eqnarray}
&&[(A_n+2)^2+(B_n+2)^2]A_{nxx}=(A_n+2)\left[A_{nx}^2-(B_{nx})^2\right]+2(B_n+2)A_{nx}B_{nx} 
\nonumber \\
&& \qquad \qquad +\frac14[(A_n+2)^2+(B_n+2)^2]\left[(A_n+B_n+4)E^2A_{n-1}+
(A_n-B_n)E^2B_{n-1}\right],\label{ABT}
\end{eqnarray}
by taking $B_n=A_n^f$ with $A_n^f$ given by \eqref{na}, \eqref{nb} or \eqref{nc}. 

The AB-Toda system \eqref{ABT} is Lax  integrable with the following Lax pair
\begin{eqnarray}
E\left(\begin{array}{c}
\phi_1 \\ \phi_2 \\ \phi_3 \\ \phi_4
\end{array}\right)
=\left(\begin{array}{cccc}
0 & 0 & 1 & 0 \\ 0 & 0 & 0 & 1 \\ -\frac12(A_n+B_n+4) & 0 & \lambda_1-u_n & 0 \\ \frac{1}{2}(B_n-A_n) & -\frac12(A_n+B_n+4) & \lambda_2 -v_n & \lambda_1-u_n
\end{array}\right)\left(\begin{array}{c}
\phi_1 \\ \phi_2 \\ \phi_3 \\ \phi_4
\end{array}\right)
\end{eqnarray}
\begin{eqnarray}
\left(\begin{array}{c}
\phi_1 \\ \phi_2 \\ \phi_3 \\ \phi_4
\end{array}\right)_t
=\left(\begin{array}{cccc}
1-\lambda_1+u_{n-1} & 0 & 1 & 0 \\ v_{n-1}-\lambda_2 & 1-\lambda_1+u_{n-1}
 & 0 & 1 \\ -\frac12(A_n+B_n+4) & 0 & 1 & 0 \\ \frac{1}{2}(B_n-A_n) & -\frac12(A_n+B_n+4) & 0 & 1
\end{array}\right)\left(\begin{array}{c}
\phi_1 \\ \phi_2 \\ \phi_3 \\ \phi_4
\end{array}\right)
\end{eqnarray}
where $u_n$ and $v_n$ are related to $A_n$ and $B_n$ by 
\begin{eqnarray}
u_{nx}=\frac12E(A_n+B_n),\ 
v_{nx}=\frac12E(A_n-B_n),
\end{eqnarray}
and $\lambda_1$ and $\lambda_2$ are two arbitrary constant spectral parameters.

\subsection{AB-$H_1$ systems} 
The same idea can also be applied to  difference-difference systems such as the $H_1,\ H_2,\ H_3,\ Q_1,\ Q_2,\ Q_3$ and $Q_4$ \cite{ABS}. Here, we discuss only some AB-$H_1$ extensions. The usual $H_1$ 
system possesses the form
\begin{equation}
(u-\hat{\tilde{u}})(\tilde{u}-\hat{u})+q-p=0,\ u=u(n,m), \ \hat{u}=u(n,m+1),\ \tilde{u} =u(n+1,m),\ \hat{\tilde{u}}=u(n+1,m+1)\label{H1}
\end{equation}
with arbitrary constants $p$ and $q$. One of its integrable coupled extension reads
\begin{eqnarray}
&&(\hat{\tilde{A}}-A)(\hat{A}-\tilde{A}+\hat{B}-\tilde{B})(\hat{A}-\tilde{A}-\hat{B}+\tilde{B})=(p_1-q_1)(\tilde{A}-\hat{A})-(p_2-q_2)(\tilde{B}-\hat{B}),\label{CH11}\\
&&(\hat{\tilde{B}}-B)(\hat{B}-\tilde{B}+\hat{A}-\tilde{A})(\hat{B}-\tilde{B}-\hat{A}+\tilde{A})=(p_1-q_1)(\tilde{B}-\hat{B})-(p_2-q_2)(\tilde{A}-\hat{A})\label{CH12}
\end{eqnarray}
with arbitrary constants $p_1,\ p_2,\ q_1$ and $q_2$.

It is straightforwardly verified that both the $H_1$ equation \eqref{H1} and the coupled $H_1$ system of equations\eqref{CH11} and \eqref{CH12} are $\hat{f}\epsilon \hat{F}$ invariant with 
\begin{equation}
F=\{\hat{P}_s^n,\ \hat{P}_s^m,\ \hat{P}_s^u,\ \hat{P}_s^n\hat{P}_s^m,\  \hat{P}_s^n\hat{P}_s^u,\ \hat{P}_s^u\hat{P}_s^m,\ \hat{P}_s^n\hat{P}_s^m \hat{P}_s^u\}
\end{equation} 
where $\hat{P}_s^u, \hat{P}_s^n$ and $\hat{P}_s^m$are defined by
\begin{equation}
\hat{P}_s^nn=-n+n_0,\ \hat{P}_s^mm=-m+m_0,\
\hat{P}_s^u\{u,\ A,\ B\}=\{-u+u_0,\ -A+u_0,\ -B+u_0\}
\end{equation}
with arbitrary constants $n_0,\ m_0$ and $u_0$. 
Because of the $\hat{P}_s^n\hat{P}_s^m$ invariance, the coupled $H_1$ system \eqref{CH11} and \eqref{CH12} allows an AB-$H_1$ reduction 
\begin{eqnarray}
&&(\hat{\tilde{A}}-A)(\hat{A}-\tilde{A}+\hat{B}-\tilde{B})(\hat{A}-\tilde{A}-\hat{B}+\tilde{B})=(p_1-q_1)(\tilde{A}-\hat{A})-(p_2-q_2)(\tilde{B}-\hat{B}),\label{ABH1}\\
&&B=A^f=\hat{f}A,\ \hat{f}\epsilon F. 
\end{eqnarray}
By fixing constants $p_2=p_1$ and $q_2=q_1$, the AB-$H_1$ system \eqref{ABH1} can be reduced to
\begin{eqnarray}
&&(\hat{\tilde{A}}-A)(\tilde{A}-\hat{A}-\hat{B}+\tilde{B})=(p_1-q_1),\label{sABH1}\\
&&B= A^f,\ \hat{f}\epsilon F.
\end{eqnarray}
The integrability of the AB-$H_1$ is integrable because of the existence of the 
Lax pair
\begin{eqnarray}
&&\hat{\phi}=\alpha\left(\begin{array}{cccc}
A & B & p_1-r_1-A\hat{A}-B\hat{B} & p_2-r_2-A\hat{B}-B\hat{A} \\
B & A &p_2-r_2-A\hat{B}-B\hat{A} & p_1-r_1 -A\hat{A}-B\hat{B}\\
1 & 0 & -\hat{A} & -\hat{B}\\
0 & 1 & -\hat{B} & -\hat{A}
\end{array}\right)\phi,\label{Ln}\\
&&\tilde{\phi}=\beta\left(\begin{array}{cccc}
A & B & p_1-r_1-A\tilde{A}-B\tilde{B} & q_2-r_2-A\tilde{B}-B\tilde{A} \\
B & A &q_2-r_2-A\tilde{B}-B\hat{A} & q_1-r_1 -A\tilde{A}-B\tilde{B}\\
1 & 0 & -\tilde{A} & -\tilde{B}\\
0 & 1 & -\tilde{B} & -\tilde{A}
\end{array}\right)\phi,\label{Lm}
\end{eqnarray}
where $\phi=(\phi_1,\ \phi_2,\ \phi_3,\ \phi_4)^T $ is the spectral function, $\alpha,\ \beta,\ r_1$ and $r_2$ are 
arbitrary constant spectral parameters. 
When $p_2=p_1$ and $q_2=q_1$, \eqref{Ln} and \eqref{Lm} becomes the Lax pair of \eqref{sABH1}. 

\section{$P_s$-$T_d$-$C$ invariant multi-soliton solutions of the AB-systems.}

In the last section, we have listed many new integrable two-place ABSs. Only two special cases, the AB-NLS system \eqref{ABNLS1} with $x_0=0$ and the complex AB-MKdV system \eqref{ABMKdV} with $x_0=t_0=0$, have been studied by some authors. 

Ablowitz and Musslimani have obtained the singular multiple soliton solutions of the nonlocal focusing equation \eqref{ABNLS1} with $x_0=0$ by using the inverse scattering transformation and Riemann-Hilbert problem \cite{AM}. Analytical dark and antidark soliton interaction solutions in the defocusing case of \eqref{ABNLS1} with $x_0=0$ have been obtained by Li and Xu \cite{PRE}. The complex soliton solutions of \eqref{ABMKdV} with $x_0=t_0=0$ have been obtained by Ji and Zhu with help of the Darboux transformations \cite{Zhu}. 

On the other hand, we know that symmetries of a nonlinear system can be used to find new solutions from known ones. The symmetry reduction method is used to find solutions which are invariant under the symmetry transformations. 

In the last section, we have established some integrable systems via $P_s$-$T_d$-$C$ invariant principles. Thus, a natural interesting question is that \em can we 
find some exact solutions via the $P_s$-$T_d$-$C$ symmetries? \rm. Especially, \em can we find $P_s$-$T_d$-$C$ invariant solutions? \rm In the following subsections, we list some $P_s$-$T_d$-$C$ group invariant solutions for general ABSs including the integrable ones proposed in the last section. 
\subsection{$P_sT_d$-invariant multi-soliton solutions of the KdV-KP-Toda type AB systems} 
Here, we discuss only the $P_s$-$T_d$-$C$ group invariant solutions, thus it is not necessary to restrict the ABSs  integrable. We call the models 
\begin{equation}
GAB\equiv G(A,\ B)=0,\quad B=A^f \label{ABG}
\end{equation} 
as the KdV-KP-Toda type ABSs if 
\begin{eqnarray}
KKT\equiv G(A,\ A)=\left\{\begin{array}{l}
KdV,\\ \\
KP,\\ \\
Toda,
\end{array}\right.\label{KKT}
\end{eqnarray}
where $KdV$, $KP$ and $Toda$ are defined by \eqref{KdV}, \eqref{KP} and \eqref{Toda}, respectively. 
Clearly, the AB-KdV, AB-KP and AB-Toda systems  listed explicitly above are all special cases of \eqref{ABG} with the condition KKT \eqref{KKT}. 

Now we focus on the $\hat{f}$-invariant solutions, i.e., 
\begin{equation}
A^f=A \label{PT}
\end{equation}
with 
$$\hat{f}=\hat{P}_s\hat{T}_d$$ 
for the KdV type equations, 
$$\hat{f}=\hat{P}^{xy}_s\hat{T}_d,\ \mbox{\rm or}\ \hat{f}=\hat{P}^{x}_s\hat{T}_d,\ \mbox{\rm or}\ \hat{f}=\hat{P}^y_s $$
 for the KP type equations, and 
$$\hat{f}=\hat{P}^{n}_s\hat{P}^x_s,\ \mbox{\rm or}\ \hat{f}=\hat{P}^x_s,\ \mbox{\rm or} \ \hat{f}=\hat{P}^n_s $$ for the Toda-type equations. In other words, to search for the $\hat{f}$ group invariant solutions of \eqref{ABG} with \eqref{KKT} is equivalent to finding the $\hat{f}$ invariant solutions of the usual KdV, KP and Toda equations. 

For the usual KdV, KP and Toda systems, various exact solutions have been found. Consequently, the only thing needs to do is to select the $\hat{f}$ invariant solutions from the known ones of the usual KdV, KP and Toda systems. Here, only the multiple soliton solutions are considered. 

It is known that for the 
KdV \eqref{KdV}, KP \eqref{KP} and Toda \eqref{Toda}, their known $N$-soliton solutions can  be uniformly written as 
\begin{eqnarray}
A&=&2\left(\ln F \right)_{xx},\label{SolA1}\\
F&=&\sum_{\mu=0,1}\exp\left(\sum_{j=1}^N\mu_j\xi_j+\sum_{1\leq j<l}^N\mu_j\mu_l \theta_{jl}\right),\label{SolF1}
\end{eqnarray}
where the summation of $\mu$ should be done for all  permutations of $\mu_i=0,\ 1, \ i=1,\ 2\ \ldots,\ N$ and 
\begin{eqnarray}
\xi_j=k_jx-k_j^3t+\xi_{0j},\quad \exp(\theta_{jl})=\left(\frac{k_j-k_l}{k_j+k_l}\right)^2 \label{theta1a}
\end{eqnarray}
for the KdV equation,
\begin{eqnarray}
\xi_j=k_jx+p_jy+k_j\left(k_j^2+\frac34p_j^2\right)t+\xi_{0j},\quad \exp(\theta_{jl})=\frac{4(k_j-k_l)^2-3(p_j-p_l)^2}{4(k_j+k_l)-3(p_j-p_l)^2}\label{theta1b}
\end{eqnarray}
for the KP equation and \begin{eqnarray}
\xi_j=k_jn+2\sinh \frac{k_j}2x+\xi_{0j},\quad \exp(\theta_{jl})=\frac{\sinh^2\frac14(k_j-k_l)}{\sinh^2\frac14(k_j+k_l)}\label{theta1c}
\end{eqnarray}
for the Toda equation.

It is clear that the multiple soliton solutions \eqref{SolA1} with \eqref{SolF1} are not explicitly $\hat{f}$-invariant. To select out the $\hat{f}$-invariant solutions from 
 \eqref{SolA1} with \eqref{SolF1}, we can use two facts,  (i) $\{\xi_{0j},\ j=1,\ 2,\ \ldots,\ n\}$ are arbitrary constants, (ii) the solution \eqref{SolA1}
 is invariant under the transformation 
\begin{eqnarray}
F\rightarrow \beta\exp(Kx+\Omega t +X_0)F \label{FF}
\end{eqnarray}
with arbitrary constants 
$\beta, \ K,\ \Omega$ and $X_0$. 

Because of the arbitrariness of $\xi_{0j}$, we rewrite $\xi_j$ as 
\begin{eqnarray}
 \xi_j&=&\eta_j-\frac12\sum_{i=1}^{j-1}\theta_{ij}-\frac12\sum_{i=j+1}^{N}\theta_{ji},\label{etaj} 
\end{eqnarray}
where 
\begin{eqnarray}
 \eta_j&\equiv &
 k_j\left(x-\frac12 x_0\right)-k_j^3\left(t-\frac12t_0\right)+\eta_{0j}\label{etajKdV}
\end{eqnarray}
for the KdV equation,
\begin{eqnarray}
\eta_j=k_j\left(x-\frac12x_0\right)+p_j\left(y-\frac12y_0\right)+k_j\left(k_j^2+\frac34p_j^2\right)\left(t-\frac12t_0\right)+\eta_{0j},\label{etajKP}
\end{eqnarray}
for the KP equation, and \begin{eqnarray}
\eta_j=k_j\left(n-\frac12n_0\right)+2\sinh \frac{k_j}2\left(x-\frac12x_0\right)+\eta_{0j},\label{etajTD}
\end{eqnarray}
for the Toda equation.

From the expressions \eqref{etajKdV}, \eqref{etajKP} and  \eqref{etajTD}, we know that 
\begin{equation}
\hat{f}\eta_j=-\eta_j+2\eta_{0j}.\label{feta}
\end{equation} 
Because of the invariance \eqref{FF}, we take the transformation 
\begin{eqnarray}
F\rightarrow \prod_{i<j}K_{ij}\exp\left(\sum_{l=1}^N \eta_l\right)F \label{FF1}
\end{eqnarray}
where 
$$K_{ij}=k_i-k_j$$
for the KdV equation,
$$K_{ij}=\sqrt{4(k_i-k_j)^2-3(p_i-p_j)^2}$$
for the KP equation 
and 
$$K_{ij}=\sinh\frac14(k_i-k_j)$$
for the Toda equation. 

By using the transformations \eqref{etaj} and \eqref{FF1}, the $N$-soliton solution \eqref{SolA1} becomes
\begin{eqnarray}
A&=&2\left[\ln \sum_{\nu=1,-1}K_\nu \cosh\left(\frac12 
\sum_{j=1}^N \nu_j\eta_j 
\right) \right]_{xx},\label{SolA1ch}
\end{eqnarray}
where the summation of $\nu=\{\nu_1,\ \nu_2,\ \ldots,\ \nu_N\}$ should be done for all non-dual permutations of $\nu_i=1,\ -1, \ i=1,\ 2\ \ldots,\ N$, and 
\begin{eqnarray}
 K_\nu=\prod_{i>j}(k_i-\nu_i\nu_jk_j), \label{theta1aa}
\end{eqnarray}
for the KdV equation,
\begin{eqnarray}
K_\nu=\prod_{i>j}\sqrt{4(k_i-\nu_i\nu_j k_j)^2-3(p_j-\nu_i\nu_jp_l)^2}\label{theta1bb}
\end{eqnarray}
for the KP equation, and 
\begin{eqnarray}
K_\nu=\prod_{i>j}\sinh\frac14(k_i-\nu_i\nu_jk_j)\label{theta1cc}
\end{eqnarray}
for the Toda equation. 

In \eqref{SolA1ch}, we have used the summation for the non-dual permutation. A pair permutations, $\nu=\{\nu_1,\ \nu_2,\ \ldots,\ \nu_N\}$ and $\nu'=\{\nu'_1,\ \nu'_2,\ \ldots,\ \nu'_N\}$, are called dual each other if 
\begin{equation}
\sum_{j=1}^N \nu_j\xi_j=-\sum_{j=1}^N \nu'_j\xi_j, \label{dual}
\end{equation} 
is true for arbitrary $\xi_j,\ j=1,\ 2,\ \ldots,\ N$. 

From the expression \eqref{SolA1ch} we know that 
\begin{equation}
A^f_{\eta_{0j}=0}=\hat{f}A_{\eta_{0j}=0}=A_{\eta_{0j}=0}.
\end{equation}
Thus the multiple soliton solution $A_{\eta_{0j}=0}$ solves all the KdV-KP-Toda type AB systems \eqref{ABG} with the conditions $\hat{f}=\hat{P}^{xy}_s\hat{T}_d$ for the AB-KP systems, and 
$\hat{f}=\hat{P}^{n}_s\hat{P}^x_s$ for the AB-Toda systems. In fact, all the other KdV-KP-Toda types systems including the whole KdV hierarchy,
 KP hierarchy, Toda hierarchy, Boussinesq hierarchy, 1+1 and 2+1 dimensional Sawada-Kortera hierarchies, 1+1 and 2+1 dimensional Kaup-Kupershmidt hierarchies, and so on, possess the same expression \eqref{SolA1ch} with suitable selections of $\eta_j$ and $K_\nu$. 

Though the expression \eqref{SolA1ch} is equivalent to \eqref{SolA1}, it appears first time in literature. For the expression \eqref{SolA1}, some types of resonant singular solutions can be obtained 
by means of the limiting procedure 
\begin{equation}
k_j\rightarrow \pm k_i. \label{kikj}
\end{equation}
However, for the expression \eqref{SolA1ch}, the direct limiting procedure of \eqref{kikj} will lead to ($N-1$) soliton solution because the transformation \eqref{FF1} moves all the possible resonant solutions to $\infty$. To recover the resonant solutions, one has to readjust the constants $\eta_{0j}$. In other words, if  letting the constants $\eta_{0j}$ depending on $k_i,\ i=1,\ 2,\ \ldots,\ N$, one may obtain quite different resonant solutions. 

\subsection{$P_sT_d$-invariant multi-soliton solutions of the MKdV-sG type AB systems}
The most general MKdV-sG system may also be written as \eqref{ABG} but with 
\begin{eqnarray}
MS\equiv G(A,\ A)=\left\{\begin{array}{l}
MKdV,\\
\\
sG,
\end{array}\right.\label{MS}
\end{eqnarray}
where $MKdV$ and $sG$ are defined by \eqref{MKdV} and \eqref{sG}, respectively. 

Using the same procedure as for the KdV-KP-Toda type AB systems, looking for the $\hat{f}=\hat{P}_s\hat{T}_d$ invariant multiple soliton solutions of the generalized MKdV-sG AB system \eqref{ABG} with \eqref{MS} is equivalent to finding the $\hat{f}$-invariant solutions of the MKdV and sG equations. 

It is known that the 
$N$ soliton solutions of the MKdV system \eqref{MKdV} and the sG equation \eqref{sG} possess the common form 
\begin{eqnarray}
A=A^{MKdV}=\frac12A^{sG}_x
=\delta i\left(\ln \frac{F_-}{F_+}\right)_x,\quad \delta^2=1
\label{SolMS}
\end{eqnarray}
with
\begin{eqnarray}
F_\pm &=&\sum_{\mu=1,0}\exp\left[\sum_{j=1}^N\mu_j\left(\xi_j\pm \frac{i\pi}2\right)+\sum_{l<j}\mu_l\mu_j \theta_{lj}\right]
\end{eqnarray}
where the summation of $\mu$ should be done for all permutations of $\mu_i=0,\ 1, \ i=1,\ 2\ \ldots,\ N$ and 
\begin{eqnarray}
\xi_j=k_jx-k_j^3t+\xi_{0j},\quad \exp(\theta_{jl})=\left(\frac{k_j-k_l}{k_j+k_l}\right)^2 \label{theta2a}
\end{eqnarray}
for the MKdV equation and 
\begin{eqnarray}
\xi_j=k_jx+k_j^{-1}t+\xi_{0j},\quad \exp(\theta_{jl})=\left(\frac{k_j-k_l}{k_j+k_l}\right)^2
\label{theta2b}
\end{eqnarray}
for the sG equation.

From the expression \eqref{SolMS}, 
we known that 
the solution is invariant under the transformation 
\begin{equation}
F_\pm \rightarrow \beta \exp(Kx+\Omega t+X_0)F_\pm  \label{Fpm}
\end{equation}
with arbitrary constants $K,\ \beta,\ \Omega,$ and $X_0$. 

Using the invariance \eqref{Fpm} and the arbitrariness of 
$\xi_{0j}$, we can rewrite the solution \eqref{Fpm} as 
\begin{eqnarray}
A&=&\left\{\begin{array}{l}
\pm 2\frac{\partial}{\partial x} \tan^{-1}\frac{\sum_{\nu_e}K_{\nu}\sinh\left(\sum_{j=1}^N \nu_j\eta_j\right)}{\sum_{\nu_o}K_{\nu}\cosh\left(\sum_{j=1}^N \nu_j\eta_j\right)},\\ \\
\pm 2\frac{\partial}{\partial x} \tan^{-1}\frac{\sum_{\nu_o}K_{\nu}\sinh\left(\sum_{j=1}^N \nu_j\eta_j\right)}{\sum_{\nu_e}K_{\nu}\cosh\left(\sum_{j=1}^N \nu_j\eta_j\right)},\\ \\
\pm 2\frac{\partial}{\partial x} \tan^{-1}\frac{\sum_{\nu_e}K_{\nu}\cosh\left(\sum_{j=1}^N \nu_j\eta_j\right)}{\sum_{\nu_o}K_{\nu}\sinh\left(\sum_{j=1}^N \nu_j\eta_j\right)},
\\ \\
\pm 2\frac{\partial}{\partial x} \tan^{-1}\frac{\sum_{\nu_o}K_{\nu}\cosh\left(\sum_{j=1}^N \nu_j\eta_j\right)}{\sum_{\nu_e}K_{\nu}\sinh\left(\sum_{j=1}^N \nu_j\eta_j\right)}
 \end{array}\right.
 \label{SolMSch}
\end{eqnarray}
via the similar trick as in the KdV-KP-Toda case, 
where 
the summation of $\nu_o$ should be done for all non-dual odd  permutations of $\nu_i=1,\ -1, \ i=1,\ 2\ \ldots,\ N$ with odd number of $\nu_i=1$, the summation of $\nu_e$ should be done for all non-dual even permutations of $\nu_i=1,\ -1, \ i=1,\ 2\ \ldots,\ N$ with even number of 
$\nu_i=1$, 
\begin{equation}
K_\nu \equiv \prod_{i>j}(k_i-\nu_i\nu_jk_j),
\end{equation}
 and $\eta_j$ being defined as 
\begin{eqnarray}
\eta_j=k_j\left(x-\frac12x_0\right)-k_j^3\left(t-\frac12t_0\right)+\eta_{0j}
\end{eqnarray} 
for the MKdV equation,
and 
\begin{eqnarray}
\eta_j=k_j\left(x-\frac12x_0\right)+k_j^{-1} \left(t-\frac12t_0\right)+\eta_{0j}
\end{eqnarray}
for the sG equation. Four expressions of \eqref{SolMSch} are equivalent by selecting suitable arbitrary constants $\eta_{0j}$ and using periodic property of the triangular functions and hyperbolic functions. 

In \eqref{SolMSch}, we have used summations on non-dual odd (even) permutations  (the permutations with odd (even) numbers of $\nu_j=1$). We should add the permutation with zero number of $\nu_j=1$ to the even permutations. For odd $N$ (for odd number of solitons), even permutations $\nu_e$ and odd permutations $\nu_o$ are dual each other, say, for $N=3$, 
\begin{eqnarray}
&&\nu_e=[\nu_1,\nu_2,\nu_3]_e=\{[-1,-1,-1],[1,1,-1],[1,-1,1],[-1,1,1]\}\nonumber \\
&&\nu_o=[\nu_1,\nu_2,\nu_3]_o=\{[1,1,1],[-1,-1,1],[-1,1,-1],[1,-1,-1]\}=\nu_e^{dual}.
\label{nu3}
\end{eqnarray}
For even $N$, $\nu_e$ and $\nu_o$ are completely different and not dual. For example, when $N=4$, the non-dual even and odd permutations possess the forms
\begin{equation}
\nu_o=[\nu_1,\nu_2,\nu_3,\nu_4]_o=\{[1,-1,-1,-1],[-1,1,-1,-1],[-1,-1,1,-1],[-1,-1,-1,1]\}, \label{nu4o}
\end{equation} 
and 
 \begin{equation}
\nu_e=[\nu_1,\nu_2,\nu_3,\nu_4]_e=\{[1,1,1,1],[1,1,-1,-1],[1,-1,1,-1],
[1,-1,-1,1]\}, \label{nu4e}
\end{equation} 
respectively. The remaining permutations are duals of \eqref{nu4o} and \eqref{nu4e}. 

From the expression \eqref{SolMSch} we know that 
\begin{equation}
A^f_{\eta_{0j}=0}=-A_{\eta_{0j}=0}.
\end{equation}
Therefore, the multiple soliton solutions $A_{\eta_{0j}=0}$ i.e., \eqref{SolMSch} with $\eta_{0j}=0$, solve the MKdV-sG AB systems \eqref{ABG} with \eqref{MS} including \eqref{ABMKdV} with $\sigma=1$ and \eqref{ABSG} with $\{a=\delta=1\}$.

\subsection{$P_sT_dC$-invariant and $P_sC$-invariant multi-soliton solutions of the NLS type AB systems}

We assume the generalized NLS-type AB systems have the following forms
\begin{equation}
G(A,\ B)=0,\qquad B=\hat{P}_s\hat{T}_d A=A(-x+x_0,-t+t_0),\ \mbox{\rm or}\ B=\hat{P}_s A=A(-x+x_0,\ t)\label{GNLS}
\end{equation}
with
\begin{equation}
G(A,\ A^*)=NLS, \label{NLSCon}
\end{equation}
and $NLS$ being defined by \eqref{NLS}. It is clear that the integrable AB-NLS systems $ABNLS_1$ and $ABNLS_2$ expressed by  \eqref{ABNLS1} and \eqref{ABNLS2} are just two explicit examples of
\eqref{GNLS}. 
 
From the condition \eqref{NLSCon}, we know that if $A$ is a $\hat{P}_s\hat{T}_d\hat{C}$ invariant solution of the usual NLS equation, then it is also a solution for all 
general NLS type AB systems \eqref{GNLS} with $\hat{f}=\hat{P}_s\hat{T}_d$.

\subsubsection{N-soliton solutions of the defocusing AB-NLS systems}
The known $N$-soliton solutions for the defocusing NLS system \eqref{NLS} with $\sigma=-1$ possesses  
the form 
\begin{eqnarray}
A=\sqrt{2}\alpha \exp(-i\alpha^2 t+i\phi_0) \frac{\sum_{\mu}\exp\left(\sum_{j=1}^{N}\mu_j(\xi_j+2i\theta_j)+\sum_{j<l}\mu_j\mu_l\theta_{jl}\right)}{\sum_{\mu}\exp\left[\sum_{j=1}^{N}\mu_j\xi_j+\sum_{j<l}\mu_j\mu_l\theta_{jl}\right]},\label{SolNLSdf}
\end{eqnarray}
where 
\begin{eqnarray}
\xi_j=
\sqrt{2}\alpha \sin(\theta_j)\left(x+\sqrt{2}\alpha \cos(\theta_j)t\right)+\xi_{0j}, \label{xijdf}
\end{eqnarray}
and
\begin{eqnarray}
&&\exp\left(\theta_{jl}\right)=\left(\frac{\sin\frac12(\theta_j-\theta_l)}{\sin\frac12(\theta_j-\theta_l)}\right)^2
\end{eqnarray}
with arbitrary real constants 
$\alpha,\ \xi_{0j},\ \theta_{j}, \ j=1,\ 2,\ \ldots,\ N $ and $\phi_0$. 
The summations on $\mu$ in the solution \eqref{SolNLSdf} should be done for all permutations $\mu_j=0,\ 1,\ j=1,\ 2,\ \ldots,\ N$. 

Similar to the KdV-KP-Toda case, the N-soliton solution \eqref{SolNLSdf} can also be re-written to an elegant form,
\begin{eqnarray}
A=\sqrt{2}\alpha \exp\left[-i\alpha^2 \left(t-\frac12t_0\right)+i\phi'_0\right] \frac{\sum_{\nu}A_\nu \cosh\left[\frac12\sum_{j=1}^{N}\nu_j\left(\eta_j+2i\theta_j\right)\right]}{\sum_{\nu}A_\nu \cosh\left(\frac12\sum_{j=1}^{N}\nu_j\eta_j\right)},\label{NLSdfch}
\end{eqnarray}
where the summations on $\nu$ should be done for all possible non-dual permutations $\nu_j=1,-1,\ j=1,\ 2,\ \ldots,\ N$,
\begin{eqnarray}
&&\eta_j=
\sqrt{2}\alpha \sin(\theta_j)\left[x-\frac12x_0+\sqrt{2}\alpha \cos(\theta_j)\left(t-\frac12t_0\right)\right]+\eta_{0j}, \label{etajdf}\\
&&A_\nu=\prod_{l<j}\sin\frac{\theta_j-\nu_j\nu_l\theta_l}2,\label{Anu}
\end{eqnarray}
and $\theta_j,\ \eta_{0j},\ \alpha$ and $\phi'_0$ are arbitrary real constants. 

The expression \eqref{NLSdfch} with \eqref{etajdf} and \eqref{Anu} is $\hat{P}_s\hat{T}_d\hat{C}$ invariant if $\phi'_0=\eta_{0j}=0$. Thus, from the condition \eqref{NLSCon} we know that 
$A_{\eta_{0j}=\phi'_0=0}$ solves all the defocusing AB-NLS systems \eqref{GNLS} with $\hat{f}
=\hat{P}_s\hat{T}_d$. 
However, for the defocusing AB-NLS \eqref{GNLS} with 
$\hat{f}
=\hat{P}_sC$, we can not 
find nontrivial multi-soliton solutions from \eqref{NLSdfch}. The fact indicates that there may exist symmetry breaking multi-soliton solutions. It is remarkable that the $\hat{P}_s$-$\hat{T}_d$-$\hat{C}$ symmetry breaking multi-soliton solutions have been obtained for the defocusing AB-NLS system \eqref{ABNLS1} with $x_0=0$ by means of the Daboux transformation \cite{PRE}. 

\subsubsection{N-soliton solutions of the focusing AB-NLS systems}

The known $N$-soliton solution for the focusing NLS system \eqref{NLS} with $\sigma=1$ possesses  
the form 
\begin{eqnarray}
A=\sqrt{2}\frac{\sum_{\mu_>}\exp\left(\sum_{j=1}^{2N}\mu_j\xi_j+\sum_{j<l}\mu_j\mu_l\theta_{jl}\right)}{\sum_{\mu_=}\exp\left(\sum_{j=1}^{2N}\mu_j\xi_j+\sum_{j<l}\mu_j\mu_l\theta_{jl}\right)},\label{SolNLS}
\end{eqnarray}
where 
\begin{eqnarray}
\xi_j=\left\{\begin{array}{ll}
k_jx+ik_j^2t+\xi_{0j},  & j=1,\ 2,\ \ldots,\ N,\\ \\
k_j^*x-ik_j^{*2}t+\xi^*_{0j},& j=N+1,\ \ldots,\ 2N,
\end{array} \right. \label{xi*}
\end{eqnarray}
\begin{eqnarray}
&&\exp\left(\theta_{jl}\right)=2(k_j-k_l)^2,\ j,l=1,\ 2,\ \ldots,\ N,
\nonumber\\ 
&&
\exp\left(\theta_{j(N+l)}\right)
=\frac1{2(k_j+k^*_l)^2},\quad 
j,l=1,\ 2,\ \ldots,\ N,\nonumber\\ &&
\theta_{(N+j)(N+l)}=\theta^*_{jl},
\end{eqnarray}
with arbitrary complex constants 
$k_{0j},\ \xi_{0j}, \ j=1,\ 2,\ \ldots,\ N. $ 

The summations on $\mu_=$ and $\mu_>$ in   the solution \eqref{SolNLS} should be done for all permutations $\mu_j=0,\ 1$ with the conditions 
\begin{equation}
\sum_{j=1}^N\mu_j=\sum_{j=1}^N\mu_{N+j},
\end{equation}
and 
\begin{equation}
\sum_{j=1}^N\mu_j=1+\sum_{j=1}^N\mu_{N+j},
\end{equation}
respectively. 

It is clear that the solution \eqref{SolNLS} is not $\hat{P}_s$-$\hat{T}_d$-$\hat{C}$ invariant for an arbitrary $\xi_{0j}$. Thus it is not a solution of the general focusing NLS type AB system \eqref{GNLS} with $\sigma=1$. 

In a similar way, to find 
$\hat{P}_s$-$\hat{T}_d$-$\hat{C}$ invariant solutions from \eqref{SolNLS}, the only thing to do is to rewrite $\xi_j$ 
as 
\begin{eqnarray}
\xi_j&=&k_j\left(x-\frac12x_0\right)+ik_j\left(t-\frac12t_0\right)+\eta_{0j}-\frac12\sum_{i=1}^{i-1}\theta_{ji}-\frac12\sum_{i=j+1}^{2n}\theta_{ij}\nonumber\\
&=& \eta_j-\frac12\sum_{i=1}^{i-1}\theta_{ji}-\frac12\sum_{i=j+1}^{2n}\theta_{ij}.\label{NLSeta} 
\end{eqnarray}

One can prove that the solution \eqref{SolNLS} with \eqref{NLSeta} is $P_sT_dC$ invariant for 
\begin{equation}
\eta_{0j}=0 \label{e0}
\end{equation}
and 
$P_sC$ invariant if
\begin{equation}
\eta_{0j}=k_j-k_j^*=0.\label{Im}
\end{equation}
Thus the solution 
$A_{\eta_{0j}=0}$ (i.e., \eqref{SolNLS} with \eqref{NLSeta} and \eqref{e0}) solves the general focusing NLS AB system \eqref{GNLS} with 
$\{B=A^{P_sT_d},\ \sigma=1\}$, 
and 
$A_{\eta_{0j}=k_j-k_k^*=0}$ (i.e., \eqref{SolNLS} with \eqref{NLSeta} 
and \eqref{Im}) solves the general focusing NLS AB system 
\eqref{GNLS} with 
$\{B=A^{{P}_s},\ \sigma=1\}$. 

The expression \eqref{SolNLS} can also be expressed in terms of hyperbolic and triangle functions to reveal explicit $\hat{P}_s\hat{T}_d\hat{C}$ invariants. However, we do dot write them down here because the final expression is much more ugly than \eqref{SolNLS}. 
 
As an example, we write down the $N=2$ case to demonstrate the explicit $\hat{P}_s\hat{T}_d\hat{C}$ invariance. In this case, the solution \eqref{SolNLS} can be rewritten as $(k_1\equiv k_{1r}+ik_{1i},\ k_2\equiv k_{2r}+ik_{2i})$
\begin{eqnarray}
A_{N=2}=\frac{2\sqrt{2}|k_1-k_2||k_1+k_{2}^*|\left[k_{1r}e^{i\eta_{1i}}\cos(\alpha+\beta+i\eta_{2r})+k_{2r}e^{i\eta_{2i}}\cos(\alpha-\beta+i\eta_{1r})\right]}{|k_1-k_2|^2\cosh(\eta_{1r}+\eta_{2r})+|k_1+k_2^*|^2\cosh(\eta_{1r}-\eta_{2r})+4k_{1r}k_{2r}\cos(\eta_{1i}-\eta_{2i}+4\beta)},\nonumber\\ \label{N=2} 
\end{eqnarray}
with
\begin{eqnarray}
&&\eta_{1i}=k_{1i}\left(x-\frac12x_0\right)+(k_{1r}^2-k_{1i}^2)
\left(t-\frac12t_0\right)+\eta_{01i}=-\frac{i}2(\eta_1-\eta_1^*),\nonumber\\
&&\eta_{2i}=k_{2i}\left(x-\frac12x_0\right)+(k_{2r}^2-k_{2i}^2)
\left(t-\frac12t_0\right)+\eta_{02i},\nonumber\\
&&\eta_{1r}=k_{1r}\left(x-\frac12x_0\right)-2k_{1r}k_{1i}\left(t-\frac12t_0\right)+\eta_{01i}=\frac12(\eta_1+\eta^*_1),\nonumber\\
&&\eta_{2r}=k_{2r}\left(x-\frac12x_0\right)-2k_{2r}k_{2i}\left(t-\frac12t_0\right)+\eta_{02i},\label{eta0i}
\end{eqnarray}
and
\begin{eqnarray}
&&\alpha=\arctan\left(\frac{k_{1i}-k_{2i}}{k_{1r}-k_{2r}}\right),\nonumber\\
&&\beta=\arctan\left(\frac{k_{1i}-k_{2i}}{k_{1r}+k_{2r}}\right).\label{abi}
\end{eqnarray} 
From \eqref{N=2} with \eqref{eta0i} and \eqref{abi}, it is straightforward to find that $A_{N=2}$ is $\hat{P}_s\hat{T}_d\hat{C}$ invariant only for $\eta_{0j}=0,\ j=1,\ 2$, and $\hat{P}_s\hat{C}$ invariant if $\eta_{0j}=k_{j}-k_j^*=0,\ j=1,\ 2$. 

\subsection{$\hat{P}_s^u\hat{P}_s^n\hat{P}_s^m$ invariant multi-soliton solutions of a special $H_1$ system} 
A general AB-$H_1$ system may have the form
\begin{equation}
G(A,\ B)=0,\ B=\hat{f}A,\ \hat{f}\epsilon \hat{F} \label{FABH1}
\end{equation}
with the condition 
\begin{equation}
G(A,\ A)=(A-\hat{\tilde{A}})(\tilde{A}-\hat{A})+q-p=0\label{AH1}
\end{equation}
for suitable $\hat{f}\epsilon \hat{F}$. 

It is clear that the special AB-$H_1$ system \eqref{sABH1} with $\sigma=1,\ q_1=2q,\ p_1=2p$ is just a special case of \eqref{FABH1} and \eqref{AH1} with  
\begin{equation}
\hat{f}=\hat{P}_s^u\hat{P}_s^n\hat{P}_s^m. 
\end{equation}
Therefore, to look for the $\hat{P}_s^u\hat{P}_s^n\hat{P}_s^m$ invariant solutions of the AB-$H_1$ equation \eqref{FABH1} is equivalent to finding the $\hat{P}_s^u\hat{P}_s^n\hat{P}_s^m$ invariant solution of the usual $H_1$ system \eqref{AH1}. 

For the usual $H_1$ equation, its $N$-soliton solution possesses the form \cite{DJZ}
\begin{equation}
A=an+bm+\gamma-\frac{h}{g}  \label{H1sola}
\end{equation} 
where $a=\sqrt{p}$ and $b=\sqrt{q}$, 
\begin{eqnarray}
\left\{
\begin{array}{l}
g=|\psi,\ E_{3}\psi, \ E_{3}^2\psi,\ \ldots,\ E_3^{N-1}\psi|, \\
 h=|\psi,\ E_{3}\psi, \ \ldots,\ E_3^{N-2}\psi,\ E_3^N\psi|,\\
\psi=\psi(n,\ m,\ l)=(\psi_1,\ \ldots,\ \psi_n)^T,\ E_3\psi=\psi(n,\ m,\ l+1),\\
\psi_i=\psi_i(n,\ m,\ l)=\rho_i^+(a+k_i)^n(b+k_i)^mk_i^l+\rho_i^-(a-k_i)^n(b-k_i)^m(-k_i)^l,
\end{array}
\right. \label{H1solb}
\end{eqnarray}
$\rho_i^{\pm}$ and $k_i$ are arbitrary constant parameters.

In the solution \eqref{H1sola} with \eqref{H1solb}, the arbitrary constants $\rho_i^+/\rho_i^-$ indicate the positions of N solitons are arbitrary and these freedoms are introduced by the translation invariance of $n$ and $m$. However, the AB-$H_1$ systems are not $\{n,\ m\}$ translation invariant. Thus, the solution \eqref{H1sola} 
with \eqref{H1solb} 
for arbitrary $\rho_i^+/\rho_i^-$ is not a solution of the AB-$H_1$ system. 

It is interesting that 
if we rewrite the constants 
$\rho_i^+/\rho_i^-$ as
\begin{equation}
\frac{\rho_i^+}{\rho_i^-}=\left(\frac{a-k_i}{a+k_i}\right)^{\frac{n_0}2}\left(\frac{b-k_i}{b+k_i}\right)^{\frac{m_0}2}e^{\xi_{0i}}
\end{equation}
with arbitrary $\xi_{i0}$,
the solution \eqref{H1sola} can be rewritten in an elegant form 
\begin{eqnarray}
&&A=an+bm+\gamma -\left[\ln(g)\right]_n=an+bm+\gamma -\frac{g_n}{g}\nonumber\\
&&=an+bm+\gamma -\frac{\sum_{\nu}K_{\nu}\sum_{j=1}^N\nu_jk_j\sinh\left(\sum_{i=1}^N\frac12\nu_i\xi_i\right)}{\sum_{\nu}K_{\nu}\cosh\left(\sum_{i=1}^N\frac12\nu_i\xi_i\right)}\label{solH1}
\end{eqnarray}
where the summation of $\nu$ should be done for all non-dual permutations of $\nu_i=1,\ -1,\ i=1,\ 2,\ \ldots,\ N$ while the $\sinh$ and $\cosh$ functions are defined by 
\begin{eqnarray}
&&\sinh \frac{\xi_i}2 =
\left(\frac{a+k_i}{a-k_i}\right)^{\frac{n}{2}
-\frac{n_0}{4}}\left(\frac{b+k_i}{b-k_i}\right)^{\frac{m}{2}
-\frac{m_0}{4}}e^{\frac{\xi_{i0}}{2}}-\left(\frac{a+k_i}{a-k_i}\right)^{\frac{n_0}{4}-\frac{n}{2}
}\left(\frac{b+k_i}{b-k_i}\right)^{\frac{m_0}{4}-\frac{m}{2}
}e^{\frac{-\xi_{i0}}{2}},\nonumber\\
&&\cosh \frac{\xi_i}2 =
\left(\frac{a+k_i}{a-k_i}\right)^{\frac{n}{2}
-\frac{n_0}{4}}\left(\frac{b+k_i}{b-k_i}\right)^{\frac{m}{2}
-\frac{m_0}{4}}e^{\frac{\xi_{i0}}{2}}+\left(\frac{a+k_i}{a-k_i}\right)^{\frac{n_0}{4}-\frac{n}{2}
}\left(\frac{b+k_i}{b-k_i}\right)^{\frac{m_0}{4}-\frac{m}{2}
}e^{\frac{-\xi_{i0}}{2}},\nonumber\\
&&\sinh(\xi_1+ \xi_2)=\sinh(\xi_1)\cosh(\xi_2)+\sinh(\xi_2)\cosh(\xi_1),\nonumber\\
&&\cosh(\xi_1+ \xi_2)=\cosh(\xi_1)\cosh(\xi_2)+\sinh(\xi_2)\sinh(\xi_1).
\end{eqnarray}
Now, it is straightforward to find that the solution \eqref{solH1} is $\hat{P}_s^u\hat{P}_s^n\hat{P}_s^m$ 
invariant if 
$\xi_{i0}=0$ for all $i=1,\ 2,\ \ldots,\ N$. In other words,
$A_{\xi_{i0}=0}$ expressed by \eqref{solH1} solves the AB-$H_1$
systems \eqref{FABH1} (including \eqref{sABH1} for $p_1=2p,\ q=2q$) with $\hat{f}=\hat{P}_s^u\hat{P}_s^n\hat{P}_s^m$.

\section{Conclusions and discussions}
In summary, various  intrinsic two-place models, named Alice-Bob systems for convenience, are proposed via the AB-BA equivalence principle and $\hat{P}_s$-$\hat{T}_d$-$\hat{C}$ principle. All the ABSs are appeared first time in this paper except for two special cases, the AB-NLS \eqref{ABNLS1} with $x_0=0$ and the AB-MKdV \eqref{ABMKdV} with
$x_0=t_0=0$. 

Though the concept of the ABSs is first proposed, all the concrete ABSs listed in this paper are 
reductions of known coupled systems. Here, many types of integrable systems such as the KdV, MKdV, KP, Toda, sG, NLS and $H_1$ equations are extended to some types of ABSs. However, in principle, any one place physical model can be extended to ABS no matter whether it is integrable or not. 

The $\hat{P}$-$\hat{T}$-$\hat{C}$ 
principles are important in many physical fields and various physically important models possess P-T-C invariance. In this paper, the parity 
$\hat{P}$ and time reversal $\hat{T}$ are generalized to shifted parity 
$\hat{P}_s$ and delayed time reversal 
$\hat{T}_d$. Thus, the 
$\hat{P}$-$\hat{T}$-$\hat{C}$ principles are extended to the
$\hat{P}_s$-$\hat{T}_d$-$\hat{C}$ principles. 

The $\hat{P}_s$-$\hat{T}_d$-$\hat{C}$ principles are used not only to establish ABSs but also to solve ABSs. 
Especially, some types of multi-soliton solutions of many ABSs have been obtained from well-known integrable models via the $\hat{P}_s$-$\hat{T}_d$-$\hat{C}$ principles. The $\hat{P}_s$-$\hat{T}_d$-$\hat{C}$ invariant multiple soliton solutions are valid also for many nonintegrable ABSs. For instance, the AB-KdV system \eqref{AAKbs}, is not integrable for general $a_i$, however, its multiple soliton solution can be obtained in terms of \eqref{SolA1ch} via a simple scaling procedure.

In addition to the $\hat{P}_s$-$\hat{T}_d$-$\hat{C}$ invariant multiple soliton solutions of many ABSs, more solutions can be obtained to be $\hat{P}_s$-$\hat{T}_d$-$\hat{C}$ symmetry breaking. For instance, for the defocusing AB-NLS system \eqref{ABNLS1} with $\sigma=-1$ and $x_0=0$, the $\hat{P}$-invariant solutions of the usual defocusing NLS are also the solutions of the defocusing $ABNLS_1$. Though we have not yet found the $\hat{P}$ invariant solution of $ABNLS_1$, there really exist other types of multiple soliton solutions which are not 
$\hat{P}$-invariant \cite{AM,PRE}. 

In \cite{PRE}, Li and Xu have obtained the explicit form of non-parity invariant multiple soliton solutions of  the defocusing $ABAKNS_1$ with $\sigma=-2$ (where 2 can be scaled to 1) and $x_0=0$ by means of the Darboux transformation, say, the two soliton solution possesses the form \cite{PRE}
\begin{equation}
A=|\kappa|e^{2i|\kappa|^2t+i\phi_0}\left\{ 1-
\frac{2\kappa_{1r}\left(\gamma
+e^{-2\kappa_{i}(x+2\kappa_{r}t)}\right)\left(\kappa^*\gamma^*
+\kappa e^{2\kappa_i(x-2\kappa_r t)}\right)}
{|\kappa|^2|\gamma|^2
+|\kappa|^2e^{-8\kappa_i\kappa_rt}
+\kappa_r\left(\kappa\gamma e^{2\kappa_i(x-2\kappa_rt)}
+\kappa^*
\gamma^*
e^{-2\kappa_i(x
+2\kappa_rt)}\right)}
\right\},\label{DNLS2}
\end{equation}
with complex constants $\kappa\equiv \kappa_r+i\kappa_i$, $\gamma$ and the real constant $\phi_0$. 

Thus, the following important question should be studied in future: \em Are there any other types of multiple soliton solutions which may be $P_s$-$T_d$-$C$ invariant or $P_s$-$T_d$-$C$ symmetry breaking for the AB systems listed in this paper? \rm

Another important question which should be studied further is \em can we establish other types of AB systems with and without AB-BA equivalence principle? \rm The answer is YES. Here, we just write down a special AB-sG system by means of the 
AB-coordinate \eqref{2d3},
\begin{eqnarray}
A_{xx}-A_{yy}=2\sin \frac{A}2\cos\frac{A^f}2,\label{ASG}
\end{eqnarray}
with 
\begin{eqnarray}
&&A^f=A(x',\ y'),\nonumber\\ \\
&&x'=\coth(\theta) x+\mbox{\rm csch}(\theta) y +\exp\left(-\frac12 \theta\right) x_0,\nonumber\\
&& y'=-\mbox{\rm csch}(\theta)x-\coth(\theta) y-\exp\left(\frac12 \theta\right)x_0.
\end{eqnarray}
The AB-sG equation \eqref{ASG} is also integrable. The Lax pair of \eqref{ASG} possesses the form 
\begin{eqnarray}
\left(\begin{array}{c}
\psi_1 \\ \psi_2 \\ \psi_3 \\\psi_4
\end{array}\right)_x =
\left(\begin{array}{cccc}
-\lambda-C_C & -iS_S & A_{x_1}+S_C & -iA^f_{x_1}-iC_S \\
iS_S & -\lambda-C_C & {i}A^f_{x_1}+{iC_S} &
+A_{x_1}+{S_C} \\
{S_C}-A_{x_1} & {i}A^f_{x_1}-{iC_S} & {\lambda}+{C_C} & 
{iS_S}\\
{iC_S}-{i}A^f_{x_1} & {S_C}
-A_{x_1} & -{iS_S} & 
{\lambda}+{C_C}
\end{array}\right)\left(\begin{array}{c}
\psi_1 \\ \psi_2 \\ \psi_3 \\\psi_4
\end{array}\right),
\end{eqnarray}
and 
\begin{eqnarray}
\left(\begin{array}{c}
\psi_1 \\ \psi_2 \\ \psi_3 \\\psi_4
\end{array}\right)_y =
\left(\begin{array}{cccc}
-\lambda+C_C & iS_S & A_{x_1}-S_C & -iA^f_{x_1}+iC_S \\
-iS_S & -\lambda+C_C & {i}A^f_{x_1}-{iC_S} &
A_{x_1}-{S_C} \\
-{S_C}-A_{x_1} & {i}A^f_{x_1}+{iC_S} & {\lambda}-{C_C} & 
-{iS_S}\\
-{iC_S}-{i}A^f_{x_1} & -{S_C}
-A_{x_1} & {iS_S} & 
{\lambda}-{C_C}
\end{array}\right)\left(\begin{array}{c}
\psi_1 \\ \psi_2 \\ \psi_3 \\\psi_4
\end{array}\right),
\end{eqnarray}
where 
\begin{eqnarray}
&&C_C=\frac{1}{16\lambda}\cos\frac{A}{2}\cos\frac{A^f}{2},\quad
S_S=\frac{1}{16\lambda}\sin\frac{A}{2}\sin\frac{A^f}{2},\nonumber\\
&&C_S=\frac{1}{16\lambda}\cos\frac{A}{2}\sin\frac{A^f}{2},\quad
S_C=\frac{1}{16\lambda}\sin\frac{A}{2}\cos\frac{A^f}{2},\nonumber\\
&&A_{x_1}=\frac18(A_x+A_y),
\quad A^f_{x_1}=\frac18(A^f_x+A^f_y),
\end{eqnarray}
and $\lambda$ is a constant spectral parameter. 

We do not discuss the model \eqref{ASG} further and leave the studies on the ABSs with different types of AB coordinates such as \eqref{2d3}, \eqref{3d3} and \eqref{3d4} in future.

\bf Acknowledgement. \rm The authors is in debt to Profs. J. F. He, Z. N. Zhu, Y. Chen, J. Lin, D. J. Zhang, X. B. Hu, Q. P. Liu, Y. Q. Li and X. Y. Tang for their helpful discussions. The work was sponsored by the Global Change Research
Program of China (No.2015CB953904), the National Natural Science Foundations of China (Nos. 11435005, 11175092, and 11205092), Shanghai Knowledge Service Platform for Trustworthy Internet of Things (No. ZF1213) and K. C. Wong Magna Fund in Ningbo University.


\end{document}